\documentstyle{article}\begin{document}
\title{Relativistic diffusive transport }
\author{ Z. Haba\\
Institute of Theoretical Physics, University of Wroclaw,\\ 50-204
Wroclaw, Plac Maxa Borna 9,
Poland}
\date{\today}\maketitle
\begin{abstract}
We discuss transport equations resulting from relativistic
diffusions in the proper time. We show that a solution of the
transport equation can be obtained from the solution of the
diffusion equation by means of an integration over the proper
time. We study the stochastic processes solving the relativistic
diffusion equation and the relativistic transport equation. We
show that the relativistic transport equation for massive
particles in the light-cone coordinates and for the massless
particles in spatial momentum coordinates are related to the
(generalized) Bessel diffusion which has an analytic solution. The
solution describes a  particle moving in a fixed direction whose
frequency distribution is the Bessel process. An approach to an
equilibrium in a moving frame is discussed. We formulate the
equilibrating diffusion and transport processes in a Lorentz
covariant way.
\end{abstract}
 \section{Introduction}
 There are many versions of the relativistic diffusion
 and of the relativistic transport (see \cite{hanggirev}\cite{debbaschrev}\cite{chev} for
 reviews).
 However, if we assume that the diffusing particle  does not change its mass,
 the diffusion evolves  in the proper
 time and that it is generated by the second order Lorentz invariant differential
 operator,
then its definition is unique as shown by Schay \cite{schay} and
Dudley \cite{dudley}. The theory can be generalized to describe  a
particle diffusing on a manifold (pseudoriemannian background
metric) \cite{lejan}\cite{franchi}. We have extended these models to
a diffusion of a particle with spin \cite{habajpa}. We have shown
\cite{habagen} that in the models of the relativistic diffusion the
energy and the angular momentum grow exponentially fast in the
proper time. Such a growth is rather unphysical in a steady world
around us. In refs.\cite{habapre}\cite{habagen} we have studied a
modification of the model by an addition of some drag terms
(friction) resulting from the requirement that a definite
equilibrium state is achieved at a large time. In \cite{habakomp} we
have shown that  a relativistic diffusion equilibrating to the
Bose-Einstein equilibrium is a linearized version of the Kompaneets
equation \cite{komp} applied for a long time to a description of
photons diffusing in an electron gas \cite{rybicki}. There are
phenomena in astrophysics \cite{rybicki}
\cite{peebles}\cite{dodelson} as well as in particle physics
\cite{ion}\cite{svet} which can be described by relativistic kinetic
and diffusion equations. The general theory of relativistic kinetic
equations has been developed in \cite{stewart}\cite{israel}. Various
forms of relativistic kinetic equations can be derived from
relativistic quantum field theory \cite{zachar}\cite{elze}
\cite{kelly}. These kinetic equations are applied to the quark-gluon
plasma \cite{svet} and to the heavy ion collisions \cite{ion} often
in the diffusion approximation.

In our earlier papers \cite{habapre}\cite{habajpa} we have
concentrated on the diffusion in the proper time. The proper time
is a reasonable mathematical tool (see \cite{landau} for its use
in relativistic dynamics) to keep explicit Lorentz invariance.
However, we should be able to describe the relativistic phenomena
also in the coordinate time (we call it the laboratory time). In
relativistic dynamics there is an equation relating the proper
time with the coordinate time and we can exchange one into the
other. The probability density as a function of the proper time
trajectories should be independent of the proper time. This
requirement is equivalent to an evolution equation of the
probability density in the coordinate time (the kinetic equation).
It is natural to interpret an analogous requirement (the diffusion
probability distribution defined on the Minkowski space-time is
independent of the proper time) as the kinetic equation (the
transport equation) in the stochastic case.

 We can obtain the transport equation by a random time change as well.
 This  is a
 stochastic analogue of the change of the proper time
 (associated with a moving particle) into a laboratory time.
 In this paper the random time change (which is a standard notion
 in the theory of diffusion processes \cite{skorohod})
is formulated as a calculational tool useful for analytic as well as
numerical calculations. The method involves an integration over the
proper time  analogous to Feynman's proper time formulation of
quantum mechanics \cite{feynman}.

In the case of  particles of moderate velocity (non-zero mass) the
meaning of the proper time diffusion can be elaborated by means of
an expansion of momenta in  powers of $(mc)^{-1}$. In this paper
we study in detail the ultrarelativistic case of massless
particles (then the proper time is a formal affine parameter on
the trajectory) and concentrated beams of massive particles when
only light-cone momenta $p_{+}$ are relevant. We perform
probabilistic calculations of the integral over the proper time
 by an application of some results of
Yor \cite{yor}\cite{yor2}. It comes out that the solution of the
transport equation obtained as an integral over the proper time
from the solution of the relativistic diffusion is expressed by
Bessel functions. This is not accidental as the transport equation
coincides with the Bessel diffusion. We compare our probabilistic
proper time method of the solution of the transport equation with
the theory of Bessel diffusions.

In general, the relativistic diffusion will have no limit for a
large proper time (no equilibrium). For the same reason the
solution of the transport equation will have no limit as the
laboratory time tends to infinity. The diffusion equation must be
supplemented by some drag terms describing a friction if the
diffusion is to achieve an equilibrium.  The equilibrium
distribution depends on the velocity of the frame where it is
observed. So, the equilibrium  is a covariant but not an invariant
notion. In this paper we formulate the dependence of the
relativistic dynamics on the equilibration reference frame in a
covariant way.

The plan of the paper is the following. In sec.2 we review the
mathematical scheme of the relativistic diffusion theory in the
proper time. In this section we also formulate and prove the basic
theorem which allows to obtain a solution of the transport
equation if the diffusion in the proper time is known. In sec.3 we
discuss various coordinates which will be applied in subsequent
sections. In sec.4 transport equations for relativistic diffusions
with drifts leading to an equilibrium are discussed. In sec.5 the
relativistic diffusion process as a function of the Brownian
motion is  obtained in an explicit form. In sec.6 a diffusion of
massless particles is discussed. Sec.7 is the main part of this
paper. We study a relation between the relativistic diffusion in
the proper time and the transport equation in the
ultrarelativistic case of massless particles and the light-cone
variables for massive particles. We solve the transport equation
by an application of the theorem of sec.2. These equations are
solved also by another method. The result is that the diffusing
particle has the momentum whose probability distribution is the
Bessel process. In sec. 8 we discuss the frame dependence of the
equilibrating diffusion process. We formulate the theory in terms
of relativistic invariant variables.
 The results are summarized  in the last section of
this paper.

\section{The general scheme of the relativistic diffusion and
relativistic transport} Following
\cite{schay}\cite{dudley}\cite{habapre} we consider an evolution in
the proper time of points on the phase space $(x,p)$
\begin{equation}
\frac{dx^{\mu}}{d\tau}=\frac{1}{m}p^{\mu},
\end{equation}
\begin{equation}
\frac{dp^{\mu}}{d\tau}=F^{\mu}
\end{equation}
preserving the mass-shell ${\cal H}_{+}$
\begin{equation}
\eta^{\mu\nu}p_{\nu}p_{\mu}=p^{\mu}p_{\mu}=p_{0}^{2}-{\bf
p}^{2}=m^{2}c^{2}.
\end{equation}
In eq.(2) $F$ can be a random force leading to a diffusive
behaviour of a test particle described by coordinates $(x,p)$.
From eq.(1) and (3) (for $m>0$) it follows that $\tau$ really has
the meaning of the proper time. Then, the evolution of a scalar
function of the trajectory is determined by the equation
\begin{equation}
\partial_{\tau}\phi={\cal G}\phi=(p^{\mu}\partial_{\mu}^{x}+{\cal
A})\phi,
\end{equation}
where (in the case of a diffusion) ${\cal A}$ is an $O(3,1)$
invariant second order differential operator; differentiation over
space-time coordinates has an index $x$ whereas differentiation
without an index concerns momenta. The probability density $\Phi$
evolves according to an adjoint equation
\begin{equation}
\partial_{\tau}\Phi_{\tau}={\cal
G}^{*}\Phi_{\tau}=(-p^{\mu}\partial_{\mu}^{x}+{\cal
A}^{*})\Phi_{\tau}
\end{equation} resulting from ($x\in R^{4}$, ${\bf p}\in R^{3}$;
fourvectors have Greek indices and  threevectors Latin indices)
\begin{equation}
\int dxd{\bf p}\phi_{\tau}(x,{\bf p})\Phi(x,{\bf p})= \int dxd{\bf
p} \phi(x,{\bf p})\Phi_{\tau}(x,{\bf p}). \end{equation}  ${\cal
A}$ is the second order differential operator such that ${\cal
A}1=0$ and the  quadratic term is negatively definite. Then, there
exists a diffusion process $(x_{\tau},p_{\tau})$ starting from
$(x,p)$ at $\tau=0$ such that
\begin{equation}
\phi_{\tau}=E[\phi(x_{\tau},p_{\tau})]\equiv{\cal K}_{\tau}\phi,
\end{equation}
where
\begin{equation}
{\cal K}_{\tau}(x,{\bf p};x^{\prime},{\bf p}^{\prime})=E[\delta(
x_{\tau}(x,p)-x^{\prime})\delta({\bf p}_{\tau}(x,{\bf p})-{\bf
p}^{\prime})].
\end{equation}
From eq.(6) \begin{equation}
 \Phi_{\tau}={\cal K}_{\tau}^{*}\Phi
\end{equation}
where
\begin{equation}
{\cal K}_{\tau}^{*}(x,{\bf p};x^{\prime},{\bf
p}^{\prime})=E[\delta(
x_{\tau}(x^{\prime},p^{\prime})-x)\delta({\bf
p}_{\tau}(x^{\prime},{\bf p}^{\prime})-{\bf p})].
\end{equation}
 The probability distribution $\Phi$ is
independent of $\tau$ if it satisfies the transport equation
\begin{equation}
{\cal G}^{*}\Phi=0.
\end{equation}

We define an equilibrium distribution $\Phi_{E}$ as an $x^{0}$
independent solution of the transport equation. Let
\begin{equation}
\Phi=\Psi\Phi_{E}.
\end{equation}
Then, $\Psi$ satisfies a diffusion equation \begin{equation}
\partial_{\tau}\Psi=\hat{{\cal G}}\Psi,
\end{equation}where
\begin{equation}
\hat{{\cal G}}=-p^{\mu}\partial_{\mu}^{x}+\hat{{\cal A}}
\end{equation}and $\hat{{\cal A}}=\Phi_{E}^{-1}{\cal A}\Phi_{E}$ is a second order differential operator
closely related to ${\cal A}$. The solution of eq.(13) can be
expressed by a stochastic process
\begin{equation}
\Psi_{\tau}=E[\Psi(\hat{x}_{\tau},\hat{p}_{\tau})].
\end{equation}

Then, if $\Phi$ is a solution of the transport equation (11) then
$\Psi$ is a solution of a diffusion equation ( we set
$t=\frac{x^{0}}{c}\geq 0$)
\begin{equation}
\partial_{t}\Psi=\frac{c}{p_{0}}{\bf p}\nabla_{\bf x} \Psi+\frac{c}{p_{0}}\hat{{\cal
A}}\Psi.
\end{equation}
 A solution of the diffusion
equation (16) can again be expressed by a diffusion process
$(\tilde{{\bf x}}_{t},\tilde{{\bf p}_{t}})$
\begin{equation}
\Psi_{t}=E[\Psi(\tilde{{\bf x}}_{t},\tilde{{\bf p}_{t}})].
\end{equation}
We can establish a relation between the solutions of the diffusion
equations (13) and (16)

{\bf Theorem}

Let $(\hat{x}_{\tau},\hat{p}_{\tau})$ be the diffusion process
(15) solving the proper time diffusion equation (13) then the
solution of the transport equation (16) with the initial condition
$\Psi({\bf x},{\bf p})$ reads
\begin{equation}
\Psi_{t}({\bf x},{\bf p})=\frac{1}{mc}\int_{0}^{\infty}d\tau
E\Big[\Psi(\hat{{\bf x}}_{\tau},\hat{{\bf p}}_{\tau})
\hat{p}_{0}(\tau)\delta\Big(t-\frac{1}{mc}\int_{0}^{\tau}ds\hat{p}_{0}(s)\Big)\Big]
\end{equation}
where \begin{displaymath}
\hat{p}_{0}(s)=\sqrt{m^{2}c^{2}+\hat{{\bf
p}}_{s}^{2}}.\end{displaymath}

{\bf Proof}: Let us first check the initial condition. For small
$t$ the proper time $\tau$ is also small and
$\int_{0}^{\tau}p_{0}(s)ds\simeq p_{0}\tau$, then $x_{\tau}\simeq
x$ and  $p_{\tau}\simeq p$. An integration over $\tau$ gives
$\Psi({\bf x},{\bf p})$ when $t\rightarrow 0$. Next, for a
derivation of the diffusion equation (16) we apply the formula
$\delta(f(\tau))=\vert f^{\prime}\vert^{-1}\delta(\tau-\tau(t))$
in order to express the time $\tau$ in the diffusion process
$({\bf x}_{\tau},{\bf p}_{\tau})$ by $t$. Then, $\vert
f^{\prime}\vert^{-1}=p_{0}(\tau)^{-1}$ cancels the $p_{0}$ term in
eq.(18).  There remains a function $\Psi$ of the process $({\bf
x}_{\tau},{\bf p}_{\tau})$ at the time $\tau(t)$ in eq.(18).This
is the random time change well-known from the theory of diffusion
processes. It is proved in \cite{skorohod} that after the random
change of time eq.(16) is satisfied (we shall still discuss the
random time change in sec.7,eqs.(76)-(79)).

 We shall apply the
formula (18) for an explicit calculation of expectation values.
Although in general the formula (18) may be difficult to use for an
analytic treatment it still  can be very useful for numerical
calculations.

We can express the solution of the transport equation (16) in
terms of a kernel $\hat{{\cal K}}_{t}$
\begin{equation} \Psi_{t}=\tilde{{\cal K}}_{t}\Psi.
\end{equation}
Then, from eqs.(17)-(18) the kernel can be expressed in the form
\begin{equation}\begin{array}{l}
\tilde{{\cal K}}_{t}({\bf x},{\bf p};{\bf x}^{\prime},{\bf
p}^{\prime})\cr =\frac{1}{mc}\int_{0}^{\infty}d\tau
E[\delta(\hat{{\bf x}}_{\tau}({\bf x},{\bf p})-{\bf
x}^{\prime})\delta(\hat{{\bf p}}_{\tau}({\bf p})-{\bf
p}^{\prime}))
p_{0}(\tau)\delta(t-\frac{1}{mc}\int_{0}^{\tau}ds\hat{p}_{0}(s))]
\cr=E[\delta(\tilde{{\bf x}}_{t}({\bf x},{\bf p})-{\bf
x}^{\prime})\delta(\tilde{{\bf p}}_{t}({\bf p})-{\bf
p}^{\prime})].\end{array}\end{equation}

In this section we have applied a notation suggesting the choice
of spatial momenta ${\bf p}$ as coordinates on the mass-shell (3).
We shall apply some other coordinates. The  definition of the
adjoint in eq.(6) is with respect to the Lebesgue measure which is
not relativistic invariant. A density factor ensuring the
invariance is contained in $\Phi$. In the next sections we shall
apply light-cone  coordinates. Then, the evolution in $x_{0}$ is
replaced by an evolution in $x_{-}=x^{0}-x^{3}$ but the
integration in eq.(6) is still with respect to a Lebesgue measure.
\section{Various coordinate systems} The relativistic diffusion is generated by the
Laplace-Beltrami operator on the mass-shell $p^{2}=m^{2}c^{2}$.
 If
we choose the spatial momenta ${\bf p}$ as coordinates on the
mass-shell then the diffusion generator reads
\begin{equation}\begin{array}{l}
2\gamma^{-2}{\cal A}=\triangle_{H}=
(\delta_{jk}+m^{-2}c^{-2}p_{j}p_{k})\frac{\partial}{\partial
p_{j}}\frac{\partial}{\partial
p_{k}}+3m^{-2}c^{-2}p_{k}\frac{\partial}{\partial p_{k}}.
\end{array}\end{equation}
We shall also use $\kappa^{2}=m^{-2}c^{-2}\gamma^{2}$ as the
diffusion constant.

The spatial momenta are convenient for a physical interpretation.
However, for a derivation of explicit solutions some other
coordinates are more useful. Let us consider the Poincare
coordinates $(q_{1},q_{2},q_{3})$ on ${\cal H}_{+}$ which are
related to momenta $p$ as follows
\begin{equation}
p_{3}+p_{0}=\frac{mc}{q_{3}},
\end{equation}\begin{displaymath}
p_{3}-p_{0}=-\frac{mc}{q_{3}}(q_{1}^{2}+q_{2}^{2}+q_{3}^{2}),
\end{displaymath}\begin{equation}
p_{1}=\frac{mc q_{1}}{q_{3}},
\end{equation}\begin{displaymath}
p_{2}=\frac{mc q_{2}}{q_{3}},
\end{displaymath}
where $q_{3}\geq 0$. Then, the metric is
\begin{equation}
ds^{2}=(mc)^{2}q_{3}^{-2}
 (dq_{1}^{2}+dq_{2}^{2}+dq_{3}^{2})
\end{equation}
and
\begin{equation}(mc)^{2}\triangle_{H}=q_{3}^{2}(\partial_{1}^{2}+\partial_{2}^{2}+\partial_{3}^{2})
-q_{3}\partial_{3}.
\end{equation}

The Poincare coordinates are closely related to the light-cone
coordinates $(p_{+},p_{a})$ (where $a=1,2$)
\begin{equation}
p_{+}=p_{0}+p_{3}.
\end{equation}
We have $q_{a}=p_{a}p_{+}^{-1}$ and $q_{3}=mcp_{+}^{-1}$. Then
\begin{equation}\begin{array}{l}
\triangle_{H}=\partial_{1}^{2}+\partial_{2}^{2}+(mc)^{-2}p_{+}^{2}\partial_{+}^{2}
+(mc)^{-2}p_{1}^{2}\partial_{1}^{2}\cr+
(mc)^{-2}p_{2}^{2}\partial_{2}^{2}+2(mc)^{-2}p_{+}p_{a}\partial_{a}\partial_{+}
+ 2(mc)^{-2}p_{1}p_{2}\partial_{1}\partial_{2} \cr
+(mc)^{-2}3p_{+}\partial_{+}+(mc)^{-2}3p_{a}\partial_{a}\end{array}\end{equation}
where $\partial_{a}=\frac{\partial}{\partial p_{a}}$. The generator
${\cal G}$ in the light-cone coordinates is
\begin{equation}
{\cal
G}=p_{-}\partial^{x}_{+}+p_{+}\partial^{x}_{-}-p_{a}\partial^{x}_{a}+\frac{\gamma^{2}}{2}\triangle_{H}.
\end{equation}
The time evolution is in $x_{-}$ instead of $x_{0}$. In the
formulae of sec.2 we replace $p_{0}$ by $p_{+}$ and $x_{0}$ by
$x_{-}$. The integration measure in eq.(6) is
$dx_{1}dx_{2}dx_{+}dx_{-}dp_{1}dp_{2}dp_{+}$.

\section{Transport equations with friction}  In order to achieve an equilibrium at a large time we add  a
friction term $K$ to the diffusion generator (21)
\begin{equation} K=K_{j}\frac{\partial}{\partial p_{j}}.
\end{equation} The transport
equation (11) with friction reads
\begin{equation}\begin{array}{l}
\frac{1}{m}\eta^{\mu\nu} p_{\nu}\frac{\partial }{\partial
x^{\mu}}\Phi\cr=
\frac{\kappa^{2}m^{2}c^{2}}{2}\frac{\partial}{\partial
p_{j}}\frac{\partial}{\partial
p_{k}}(\delta_{jk}+m^{-2}c^{-2}p_{j}p_{k})\Phi-\frac{3\kappa^{2}}{2}\frac{\partial}{\partial
p_{k}}p_{k}\Phi-\frac{\partial}{\partial p_{k}}K_{k}\Phi.
\end{array}\end{equation} Eq.(30) determines the drift $K$ if $\Phi_{E}$ is space-time independent  \begin{equation}\begin{array}{l}
K_{k}=\kappa^{2}m^{2}c^{2}\Phi_{E}^{-1}\Big(\frac{1}{2}\frac{\partial}{\partial
p_{j}}(\delta_{jk}+m^{-2}c^{-2}p_{j}p_{k})-\frac{3}{2}(mc)^{-2}p_{k}\Big)\Phi_{E}.\end{array}
\end{equation} Let us consider a class of solutions $\Phi_{E}$ depending solely
on $p_{0}$. We write $\Phi_{E}$ in the form
\begin{equation}
\Phi_{E}=p_{0}^{-1}\exp(f(\beta cp_{0})).
\end{equation}
The $p_{0}^{-1}$ factor ensures the Lorentz invariance of the
measure $d{\bf p}p_{0}^{-1}$. From eq.(31) we obtain
\begin{equation}
K_{k}=\frac{\kappa^{2}}{2}p_{k}\beta cp_{0}f^{\prime}(\beta c
p_{0}).
\end{equation}

We choose $\Phi$ in the form
\begin{equation}
\Phi=p_{0}^{-1}\exp( f(c\beta p_{0}))\Psi
\end{equation}

Then,  the transport equation (16) reads
\begin{equation}
\partial_{t}\Psi=\frac{c}{p_{0}}({\bf p}\nabla_{{\bf x}}+\hat{{\cal A}})\Psi
\end{equation}
where
\begin{equation}
\hat{{\cal A}}=K_{j}\partial^{j}+{\cal A}
\end{equation}
and ${\cal A}$ is defined in eq.(21).

Explicitly
\begin{equation}\begin{array}{l}
\frac{\partial}{\partial
x^{0}}\Psi=\frac{1}{p_{0}}p_{j}\frac{\partial}{\partial x_{j}}\Psi
 \cr
+\frac{\kappa^{2}m^{3}c^{2}}{2p_{0}}(\delta_{jk}+m^{-2}c^{-2}p_{j}p_{k})\frac{\partial}{\partial
p_{j}}\frac{\partial}{\partial
p_{k}}\Psi+\frac{m\kappa^{2}}{p_{0}}(\frac{3}{2}+\frac{1}{2}
c\beta f^{\prime}(\beta cp_{0})p_{0})p_{k}\frac{\partial}{\partial
p_{k}}\Psi.
\end{array}\end{equation}(the formulae (35)-(37) are not obvious but come out
from  calculations). The transport equation (without friction)  in
light-cone coordinates is
\begin{equation}
p_{-}\partial^{x}_{+}+p_{+}\partial^{x}_{-}\Psi-p_{a}\partial_{a}\Psi-\frac{\gamma^{2}}{2}\triangle_{H}^{*}\Psi=0
\end{equation}
where $x_{+}=x^{0}+x^{3}$ and $x_{-}=x^{0}-x^{3}$. The adjoint of
the operator (27) in eq.(38) is with respect to the Lebesgue measure
$dp_{+}dp_{1}dp_{2}$.

We look for an equilibrium solution of the transport equation in
the form (it is not normalizable with respect to the integration
over the transverse momenta $p_{a}$)
\begin{equation}
\Phi_{E}=p_{+}^{-1}\exp(f(\beta c p_{+})).
\end{equation}
We shall still discuss  equilibrium distributions of the form (39)
in sec.8. Here, we only point out that the light-cone coordinates
are appropriate for particle beams moving in the direction of the
third axis such that the remaining momenta $p_{-}$ and $p_{a}$ are
small. An analogue of eq.(31) gives
\begin{equation}\begin{array}{l}
K=K_{-}\partial_{+}+K_{a}\partial_{a}=\frac{1}{2}\beta c
\kappa^{2}p_{+}^{2}f^{\prime}\partial_{+}
+\kappa^{2}p_{+}\partial_{+}-\frac{3}{2}\kappa^{2}p_{a}\partial_{a}
\end{array}\end{equation}
for the friction leading to the equilibrium (39). A particular
(non-normalizable but relativistic invariant) distribution
corresponds to $f=0$. We note that the change of coordinates
$(p_{1},p_{2},p_{3})\rightarrow (p_{1},p_{2},p_{+})$ transforms
the relativistic invariant measure $dp_{1}dp_{2}dp_{3}p_{0}^{-1}$
into
\begin{displaymath}
dp_{1}dp_{2}dp_{+}p_{+}^{-1}.
\end{displaymath}
With the friction (40) the diffusion generator (4) reads

\begin{equation}\begin{array}{l}
2\gamma^{-2}{\cal
A}=\partial_{1}^{2}+\partial_{2}^{2}+(mc)^{-2}p_{+}^{2}\partial_{+}^{2}
+(mc)^{-2}p_{1}^{2}\partial_{1}^{2}\cr+
(mc)^{-2}p_{2}^{2}\partial_{2}^{2}+2(mc)^{-2}p_{+}p_{a}\partial_{a}\partial_{+}
+ 2(mc)^{-2}p_{1}p_{2}\partial_{1}\partial_{2} \cr
+5(mc)^{-2}p_{+}\partial_{+} + \beta c
\frac{1}{m^{2}c^{2}}p_{+}^{2}f^{\prime}\partial_{+}
\end{array}\end{equation}
and the diffusion operator $\hat{{\cal A}}$ of eqs.(13)-(14) and
eq.(36) in the evolution equation for the proper time probability
distribution
\begin{displaymath}
\partial_{\tau}\Psi=(-p_{-}\partial^{x}_{+}-p_{+}\partial^{x}_{-}\Psi+p_{a}\partial_{a}\Psi
+\hat{{\cal A}})\Psi
\end{displaymath}is
\begin{equation}\begin{array}{l}
2\gamma^{-2}\hat{{\cal
A}}=\partial_{1}^{2}+\partial_{2}^{2}+(mc)^{-2}p_{+}^{2}\partial_{+}^{2}
+(mc)^{-2}p_{1}^{2}\partial_{1}^{2}\cr+
(mc)^{-2}p_{2}^{2}\partial_{2}^{2}+2(mc)^{-2}p_{+}p_{a}\partial_{a}\partial_{+}
+ 2(mc)^{-2}p_{1}p_{2}\partial_{1}\partial_{2} \cr
+(mc)^{-2}p_{+}\partial_{+}+(mc)^{-2}2p_{a}\partial_{a} + \beta c
\frac{1}{m^{2}c^{2}}p_{+}^{2}f^{\prime}\partial_{+}+ 2\beta c
\frac{1}{m^{2}c^{2}}p_{+}p_{a}f^{\prime}\partial_{a}.
\end{array}\end{equation}
\section{Stochastic equations}

In the Poincare coordinates the diffusion process is a solution of
the linear stochastic differential equations
($\gamma^{2}=m^{2}c^{2}\kappa^{2}$)
\begin{displaymath}
dq_{a}=\kappa q_{3}db_{a},
\end{displaymath}
$a=1,2$,
\begin{equation}
dq_{3}=-\frac{\kappa^{2}}{2}q_{3}d\tau+\kappa
q_{3}db_{3}=-\kappa^{2}q_{3}d\tau+\kappa q_{3}\circ db_{3},
\end{equation}
where Stratonovitch differentials are denoted by a circle and the
Ito stochastic differentials without the circle (the notation is
the same as in \cite{ikeda}). The Brownian motion appearing on the
rhs of  eqs.(43) is defined as the Gaussian process with the
covariance
\begin{equation}
E[b_{a}(\tau)b_{c}(s)]=\delta_{ac}min(\tau,s). \end{equation}
 The solution of eq.(43) is
\begin{equation}
q_{3}(\tau)=\exp(-\kappa^{2}\tau+\kappa b_{3}(\tau))q_{3}
\end{equation}
and
\begin{equation}
q_{a}(\tau)=q_{a}+\kappa\int_{0}^{\tau}q_{3}(s)db_{a}(s).
\end{equation}

 In the light-cone coordinates
\begin{equation}
p_{\pm}=p_{0}\pm p_{3}
\end{equation}
we have
\begin{equation}
dp_{+}=\kappa^{2}p_{+}d\tau+\kappa p_{+}\circ
db_{+}=\frac{3\kappa^{2}}{2}p_{+}d\tau+\kappa p_{+} db_{+},
\end{equation} \begin{equation}
dp_{a}=\kappa^{2}p_{a}d\tau-\kappa p_{a} db_{+}+\gamma db_{a}
\end{equation}
where $a=1,2$.

The solution of eqs.(49) is
\begin{equation}\begin{array}{l}
p_{a}(\tau)=\exp(\kappa^{2}\tau-\kappa
b_{+}(\tau))p_{a}+\gamma\int\exp(\kappa^{2}s-\kappa
b_{+}(s))db_{a}(s)\equiv
e_{a}+\delta_{a}.\end{array}\end{equation} The similarity between
the solutions in light-cone coordinates and the Poincare
coordinates is not accidental. It follows from the relation (23)
between these coordinates.

A relativistic transport equation can be important in applications
to high-energy plasma.
 In an electromagnetic field there is an additional drift term in the diffusion equation
 (4)
\begin{equation}
K=\frac{e}{mc}F_{j\nu}p^{\nu}\partial^{j},
\end{equation}
where $F_{\mu\nu}$ denotes the electromagnetic field tensor.
 The
stochastic equations are linear and can be solved if the only
components of $F$ are $F_{12}=B$ and $F_{30}=E$. In such a case
the stochastic equations (48)-(49) read
\begin{equation}
dp_{1}=\kappa^{2}p_{1}d\tau+\alpha Bp_{2}d\tau-\kappa p_{1}
db_{+}+\gamma db_{1},
\end{equation}
\begin{equation}
dp_{2}=\kappa^{2}p_{2}d\tau-\alpha Bp_{1}d\tau-\kappa p_{2}
db_{+}+\gamma db_{2},
\end{equation}
\begin{equation}
dp_{+}=\kappa^{2}p_{+}d\tau+\alpha Ep_{+}d\tau +\kappa p_{+}\circ
db_{+},
\end{equation}
where
\begin{displaymath}
\alpha=\frac{e}{mc}.
\end{displaymath}
Eqs.(52)-(54) should be supplemented by equations determining the
coordinates $dx_{a}=m^{-1}p_{a}d\tau$, $dx_{+}=m^{-1}p_{-}d\tau$
and \begin{displaymath} dx_{-}=m^{-1}p_{+}d\tau.\end{displaymath}

A particular solution $\Phi_{E}$ of the transport equation in an
electric field (51) ($F_{30}=E$, the remaining $F_{\mu\nu}=0$ in
eq.(51)) \begin{displaymath}
-p_{+}\partial^{x}_{-}\Phi-p_{-}\partial^{x}_{+}\Phi+p_{a}\partial_{a}^{x}\Phi
+{\cal A}^{*}\Phi-\alpha E\partial_{+}p_{+}\Phi=0,
\end{displaymath}
where ${\cal A}^{*} $ is the adjoint of ${\cal A}$ (eq.(41) with
$f=0$), is

\begin{equation} \Phi_{E}=p_{+}^{-1}.
\end{equation}
Let $\Phi=p_{+}^{-1}\Psi$ then  the proper time diffusion equation
for $\Psi$ takes the form\begin{displaymath}
\partial_{\tau}\Psi=
(-p_{+}\partial^{x}_{-}-p_{-}\partial^{x}_{+}+p_{a}\partial_{a}^{x}
+\hat{{\cal A}}+\alpha E p_{+}\partial_{+})\Psi.
\end{displaymath}
 The transport equation reads
\begin{equation}\begin{array}{l}
\partial^{x}_{-}\Psi=-p_{+}^{-1}p_{-}\partial^{x}_{+}\Psi+p_{+}^{-1}p_{a}\partial_{a}^{x}\Psi
+p_{+}^{-1}\hat{{\cal A}}\Psi+\alpha
E\partial_{+}\Psi,\end{array}\end{equation} where $\hat{{\cal A}}$
is defined in eq.(42) with $f=0$. In the light-cone coordinates
the stochastic equation for the process $\hat{p}_{+}$ of eq.(15)
in an electric field $E$ (but without friction) is
\begin{equation}
d\hat{p}_{+}=\frac{1}{2}\kappa^{2}\hat{p}_{+}d\tau+\alpha E
\hat{p}_{+}d\tau +\kappa \hat{p}_{+}db_{+}.
\end{equation} It has the solution
\begin{equation} \hat{p}_{+}(\tau)=\exp(\alpha
E\tau+\kappa b_{+}(\tau)).
\end{equation}
If we have a friction leading to the J\"uttner equilibrium
distribution  (without an electromagnetic field and
$f^{\prime}=-1$) then the stochastic equation for $\hat{p}_{+}$ is
\begin{equation}
d\hat{p}_{+}=\frac{1}{2}\kappa^{2}\hat{p}_{+}d\tau-\frac{1}{2}\beta
c\kappa^{2}\hat{p}_{+}^{2}d\tau +\kappa \hat{p}_{+}db_{+}.
\end{equation}
We discuss its solutions in sec.7.
\section{Massless particles} There
is a substantial simplification of stochastic equations if $m=0$.
In order to obtain the limit of zero mass we let
$\delta_{jk}\rightarrow 0$ in eq.(21). The diffusion generator in
the limit $m\rightarrow 0$ reads
\begin{equation}\begin{array}{l}
\triangle_{H}= p_{j}p_{k}\frac{\partial}{\partial
p_{j}}\frac{\partial}{\partial
p_{k}}+3p_{k}\frac{\partial}{\partial p_{k}}.
\end{array}\end{equation}

 The transport equation for a diffusion with a friction is
 (when $m=0$ we use only the diffusion constant $\kappa^{2}$)
\begin{equation}\begin{array}{l}
\eta^{\mu\nu}p_{\nu}\frac{\partial }{\partial x^{\mu}}\Phi=
\frac{\kappa^{2}}{2}\frac{\partial}{\partial
p_{j}}\frac{\partial}{\partial
p_{k}}p_{j}p_{k}\Phi-\frac{3\kappa^{2}}{2}\frac{\partial}{\partial
p_{k}}p_{k}\Phi-\frac{\partial}{\partial
p_{k}}K_{k}\Phi.\end{array}\end{equation} If $\Phi_{E}$ depends
only on momenta then from eq.(61)
\begin{equation}\begin{array}{l}
K_{k}=\kappa^{2}\Phi_{E}^{-1}\Big(\frac{1}{2}\frac{\partial}{\partial
p_{j}}p_{j}p_{k}-\frac{3}{2}p_{k}\Big)\Phi_{E}.\end{array}
\end{equation}
Then,  eq.(13) reads (in the massless case the proper time is just
an affine time parameter without any physical meaning)
\begin{equation}
\begin{array}{l}
\partial_{\tau}\Psi_{\tau}=
\frac{1}{2}\kappa^{2}p_{j}p_{k}\partial^{j}\partial^{k}\Psi_{\tau}
+2\kappa^{2}p_{j}\partial^{j}\Psi_{\tau}+\frac{1}{2}\kappa^{2}p_{j}p_{k}(\partial^{j}
 \ln\Phi_{E})
\partial^{k}\Psi_{\tau}-p_{\mu}\partial_{x}^{\mu}\Psi_{\tau}.
\end{array}\end{equation}
The transport equation follows from eqs.(62)-(63). In an explicit
form
\begin{equation}\begin{array}{l}
\frac{\partial}{\partial
x^{0}}\Psi=\frac{1}{p_{0}}p_{j}\frac{\partial}{\partial x_{j}}\Psi
+\frac{\kappa^{2}}{2p_{0}}p_{j}p_{k}\frac{\partial}{\partial
p_{j}}\frac{\partial}{\partial p_{k}}\Psi\cr
+2\kappa^{2}p_{j}p_{0}^{-1}\partial^{j}\Psi
+\frac{1}{2}\kappa^{2}p_{j}p_{k}p_{0}^{-1}(\partial^{j}
 \ln\Phi_{E})
\partial^{k}\Psi.
\end{array}\end{equation}
We  discuss in more detail the J\"uttner equilibrium distribution
\cite{juttner}
\begin{equation} \Phi_{E}=p_{0}^{-1}\exp(-c\beta p_{0}).
\end{equation}Then, from eq.(62) we obtain
\begin{equation}
K_{k}=-\frac{c\kappa^{2}}{2}\beta p_{k}p_{0}.
\end{equation}
The stochastic equation for the diffusion (15) in the proper time
with the friction (66) reads
\begin{equation}
 d\hat{p}_{j}=
 \frac{3\kappa^{2}}{2}\hat{p}_{j}d\tau
 -\frac{c\beta\kappa^{2}}{2}\hat{p}_{j}\vert\hat{{\bf p}}\vert d\tau+\kappa
 \hat{p}_{j}db.
 \end{equation}
From eq.(67) we obtain the stochastic equation for
$\hat{p}_{0}=\vert \hat{ {\bf p}}\vert$
\begin{equation}
d\hat{p}_{0}=\frac{3\kappa^{2}}{2}\hat{p}_{0}d\tau-\frac{c\kappa^{2}}{2}\beta
\hat{p}_{0}^{2}d\tau +\kappa \hat{p}_{0}db.
\end{equation}

\section{Solution of the transport equation }
In this section we wish to apply the Theorem  of sec.2 in order to
derive a solution of the transport equation. Let us begin with the
spatial momenta as coordinates on the mass-shell ${\cal H}_{+}$.
The stochastic equation for the transport diffusion $(\tilde{{\bf
x}}_{t},\tilde{{\bf p}}_{t})$ (17) of a massive particle  solving
eq.(37)  in ${\bf p}$ coordinates reads
\begin{equation}\begin{array}{l}
d\tilde{p}_{j}=\frac{\tilde{p}_{j}\kappa^{2}m}{2} c^{2}\beta
f^{\prime}dt+\frac{3\kappa^{2}mc}{2\tilde{p}_{0}}\tilde{p}_{j}dt
+mc\sqrt{mc}\tilde{p}_{0}^{-\frac{1}{2}}\kappa db_{j}\cr + \sqrt{mc}
\kappa
(\tilde{p}_{0}^{\frac{1}{2}}-mc\tilde{p}_{0}^{-\frac{1}{2}})\tilde{{\bf
p}}^{-2}\tilde{p}_{j}\tilde{{\bf p}}d{\bf b}\cr \equiv
\frac{\tilde{p}_{j}\kappa^{2}m}{2} c^{2}\beta
f^{\prime}dt+\frac{3\kappa^{2}mc}{2\tilde{p}_{0}}\tilde{p}_{j}dt
+\sqrt{c} \kappa{p}_{0}^{-\frac{1}{2}}e_{j}^{n} db_{n},
\end{array}\end{equation}
\begin{equation}
d\tilde{x}^{j}=-c\tilde{p}_{j}\tilde{p}_{0}^{-1}dt,
\end{equation}
here we defined the square root of the metric (the second order
terms in eq.(21))
\begin{displaymath}
e^{n}_{j}e^{n}_{k}=\delta^{jk}+(mc)^{-2}p_{j}p_{k}.
\end{displaymath}
 The process (69)-(70) gives the solution of
the transport equation (37) in the form
\begin{displaymath}\Psi_{t}({\bf x},{\bf p})=E[\Psi(\tilde{{\bf x}}_{t}({\bf x},{\bf p}),\tilde{{\bf
p}}_{t}({\bf p}))].
\end{displaymath}
Unfortunately, the non-linear equations (69)-(70) are difficult to
solve explicitly. We could also consider the proper time equations
in these coordinates
 discussed extensively in \cite{habapre}. Then, we could apply the Theorem
of sec.2. However, such a method would be fruitful only in numerical
calculations. Computing the expectation values (18) can be more
efficient then solving  eqs.(69)-(70) directly.

The light-cone coordinates are more useful for analytic
solutions.The formula (18) for the solution of the transport
equation in these coordinates takes the form
\begin{equation}\begin{array}{l}
\Psi(x_{-};x_{+},x_{a},p_{+},p_{a})\cr=\frac{1}{m}\int_{0}^{\infty}d\tau
E\Big[\Psi(\hat{x}_{a}(\tau),\hat{x}_{+}(\tau),\hat{p}_{+}(\tau),\hat{p}_{a}(\tau))
\delta\Big(x_{-}-\frac{1}{m}\int_{0}^{\tau}ds\hat{p}_{+}(s)ds\Big)\hat{p}_{+}(\tau)\Big],\end{array}
\end{equation}
where $\hat{p} $ is the stochastic process (15)   generated by
$\hat{{\cal A}}$ of eq.(42). If there is no friction then the
stochastic equations of the diffusion (42) are linear and can be
solved explicitly (see eqs.(50) and (58)). Then, the expectation
value of a function $h$ of $\hat{p}_{a}$ has a representation in
terms of its Fourier transform
\begin{displaymath}\begin{array}{l}E[h(\hat{p}_{1}(\tau),\hat{p}_{2}(\tau))]=\int
dk_{1}dk_{2}\tilde{h}(k_{1},k_{2})
E[\exp(ik_{a}\hat{p}_{a}(\tau))]\cr =\int
dk_{1}dk_{2}\tilde{h}(k_{1},k_{2})
E\Big[\exp(ik_{a}e_{a}(\tau))\exp\Big(-\frac{\gamma^{2}}{2}k_{a}k_{a}
\int \exp(2\gamma^{2}s+2\gamma b_{+}(s))ds\Big)\Big]\end{array}
\end{displaymath}
($e_{a}$ is defined in eq.(50)).The joint distribution law of
$b,\int \exp b,\int\exp 2b$ can be explicitly calculated
\cite{alili}. Hence, a solution of the transport equation
(including the electromagnetic field (51)) can be obtained in a
form of an integral kernel.

We consider  here a simplified version of the Theorem of sec.2
when the initial condition $\Psi$ depends solely on $p_{+}$. Let
us note that $p_{+}$ enters the generator of the diffusion with
the factor $(mc)^{-1}$ (or with $\kappa=(mc)^{-1}\gamma$). Hence,
it is irrelevant for moderate velocities. Our restriction to
probabilities depending only on $p_{+}$ in fact applies to the
ultrarelativistic limit when the transverse momenta can be
neglected. In such a case  the transport equation with the
 J\"uttner friction (39) ($f^{\prime}=-1$) but without the electric
field reads
\begin{equation}\begin{array}{l}
\partial_{-}\Psi=\frac{\kappa^{2}}{2}p_{+}\partial_{+}^{2}\Psi
+\frac{\kappa^{2}}{2}\partial_{+}\Psi -\frac{1}{2}c\beta\kappa^{2}
p_{+}\partial_{+}\Psi.
\end{array}\end{equation} The transport equation with the electric field
$E$ and without friction ($f=0$ in eq.(39)) defined in eq.(56) can
be written explicitly as

\begin{equation}\begin{array}{l}
\partial_{-}\Psi=\frac{\kappa^{2}}{2}p_{+}\partial_{+}^{2}\Psi
+(\frac{\kappa^{2}}{2}+\alpha E)\partial_{+}\Psi.
\end{array}\end{equation}

 Eq.(71) gives the solution of eqs. (72)-(73) in the form
\begin{equation}
\begin{array}{l}\Psi(x_{-},p_{+})\cr
=m^{-1}\int_{0}^{\infty}d\tau E\Big[\Psi(p_{+}(\tau,p_{+}))
\delta\Big(x_{-}-\frac{1}{m}\int_{0}^{\tau}ds p_{+}(s,p_{+})\Big)
p_{+}(\tau,p_{+})\Big].\end{array}
\end{equation}
We can solve the diffusion equations (72)-(73) by means of the
stochastic process (17). For eq.(72) the stochastic equation is
\begin{equation}
d\tilde{p}_{+}=\frac{\kappa^{2}}{2}dx_{-}-\frac{1}{2}\beta c
\kappa^{2}\tilde{p}_{+}dx_{-}+\kappa\sqrt{\tilde{p}_{+}}db
\end{equation}
whereas the stochastic process solving  eq.(73) satisfies the
equation
\begin{equation}
d\tilde{p}_{+}=(\frac{\kappa^{2}}{2}+\alpha
E)dx_{-}+\kappa\sqrt{\tilde{p}_{+}}db.
\end{equation} We shall discuss solutions of eqs.(72)
and (73) together with the
 transport equation for massless particles because
  all these equations are related to the Bessel diffusion.

Before solving the equations  let us explain the text-book method
\cite{skorohod} of the random change of time in eq.(67) (then we
show that the Theorem of sec.2 is its efficient realization). Let us
treat the formula (1) for $x^{0}(\tau)$
 as a definition of the proper time $\tau$
\begin{equation}
\tau=\int_{0}^{x^{0}}\vert \hat{{\bf p}}_{s}\vert^{-1}ds
\end{equation}
(here $\hat{{\bf p}}_{s}$ is the solution of the proper time
diffusion (67); $x^{0}(\tau)$ can be defined implicitly by (77)).
We can see from eq.(77) that $\tau $ depends only on events
earlier than $x^{0}$. As a consequence
$\hat{p}_{j}(x^{0})=p_{j}(\tau(x^{0}))$ is again a Markov process.
 Then, differentiating the momenta and coordinates according to the rules of the Ito calculus
 \cite{ikeda} we obtain the following Langevin equations (for
mathematical details of a random change of time see
\cite{skorohod}\cite{ikeda}; a random time change from $x^{0}$ to
$\tau$ is discussed in \cite{dun})
\begin{equation}\begin{array}{l}
dp_{j}(x^{0})=
 \frac{3\kappa^{2}}{2}\vert{\bf p}\vert^{-1}p_{j}dx^{0}
 -\frac{c\beta\kappa^{2}}{2}p_{j} dx^{0}+
 \kappa p_{j}\vert{\bf p}\vert^{-\frac{1}{2}}db,

\end{array}\end{equation}
\begin{equation}
dx^{j}=-p_{j}\vert {\bf p}(x^{0})\vert^{-1}dx^{0}.
\end{equation}
Let  $\Psi({\bf x},{\bf p})$ be an arbitrary function of ${\bf x}$
and ${\bf p}$ and $(x(x^{0},x,{\bf p}),{\bf p}(x^{0},{\bf p}))$ the
solution of eqs.(78)-(79) with the initial condition $({\bf x},{\bf
p})$ then
\begin{equation} \Psi_{t}({\bf x},{\bf p})=E\Big[\Psi\Big({\bf
x}(x^{0},{\bf x},{\bf p}),{\bf p}(x^{0},{\bf p})\Big)\Big]
\end{equation}
 is the solution of the transport equation
\begin{equation}\begin{array}{l}
\frac{\partial}{\partial
x^{0}}\Psi=\frac{1}{p_{0}}p_{j}\frac{\partial}{\partial x_{j}}\Psi
+\frac{\kappa^{2}}{2p_{0}}p_{j}p_{k}\frac{\partial}{\partial
p_{j}}\frac{\partial}{\partial p_{k}}\Psi-\frac{\kappa^{2}}{2}\beta
c p_{k}\frac{\partial}{\partial
p_{k}}\Psi+\frac{3\kappa^{2}}{2p_{0}} p_{k}\frac{\partial}{\partial
p_{k}}\Psi
\end{array}\end{equation} with the initial condition $\Psi({\bf
x},{\bf p})$. We can prove that  eq.(81) is satisfied by
differentiation
 of eq.(80) and application of  the rules of the Ito calculus \cite{ikeda} (or the well-known
 relation between  Langevin equation and the diffusion equation).

We return to the diffusion in the proper time in order to perform
the average over the random time $\tau(x^{0})$ as prescribed by
eq.(18) (this formula is treating the random change of time as a
computational tool). From eqs.(67)-(68) it follows that
\begin{equation}
 \hat{{\bf p}}_{\tau}\vert \hat{{\bf
p}}_{\tau}\vert^{-1}\equiv {\bf
n}=(\cos\phi\sin\theta,\sin\phi\sin\theta,\cos\theta)=const
\end{equation}
 is time independent. Hence, according to eq.(18) the solution of the
transport equation (64) takes the form
\begin{equation}
\Psi_{t}({\bf x},\vert {\bf p}\vert,{\bf
n})=\frac{1}{mc}\int_{0}^{\infty}d\tau E\Big[\Psi({\bf
x}(\tau,{\bf x}),\vert \hat{{\bf p}}_{\tau}\vert,{\bf
n})\delta\Big(t-\frac{1}{mc} \int_{0}^{\tau}\vert \hat{{\bf
p}}_{s}\vert ds\Big)\vert \hat{{\bf p}}_{\tau}\vert\Big],
\end{equation}
where
\begin{equation}
{\bf x}(\tau,{\bf x})={\bf x}-\int_{0}^{\tau}ds\vert \hat{{\bf
p}}_{s}\vert {\bf n}.
\end{equation}

In order to calculate the expectation value (83) we apply the
results of Yor \cite{yor}\cite{yor2} who obtained (using the
Feynman-Kac formula for an exponential potential) the joint
distribution of
\begin{equation}
B^{(\mu)}(\tau)=b_{\tau}+\mu \tau
\end{equation}
and
\begin{equation}
A^{(\mu)}(\tau)=\int_{0}^{\tau}ds\exp\Big(2 B^{(\mu)}(s)\Big).
\end{equation}
We need some rescalings (using $\alpha b(s)=b(\alpha^{2} s)$) in
order to bring our formulae to the Yor's form. We write
\begin{equation}
\exp(\kappa^{2}s+\kappa
b(s))=\exp\left(2\left(2\kappa^{2}\frac{s}{4}+b\left(\kappa^{2}\frac{s}{4}\right)\right)\right).
\end{equation}
It follows that $\mu=2$ in Yor's formula.

 After such an
rescaling an expectation value of a function of
 $B_{\tau}=\kappa^{2}\tau +\kappa b(\tau)$ and $A(\tau)=\int_{0}^{\tau}ds\exp
2B_{s}$ is \cite{yor}\cite{yor2}\begin{equation}\begin{array}{l}
E\Big[F\Big(B(\tau),A(\tau)\Big)\Big]=\exp(-\frac{\tau\kappa^{2}}{2})\int
F(v,\frac{4}{\kappa^{2}}\vert{\bf p}\vert y)\exp(2v) \cr
\exp\Big(-\frac{1}{2y}(1+\exp(2v))\Big)\theta(y^{-1}\exp(v),\frac{\tau\kappa^{2}}{4})y^{-1}dydv
\end{array}\end{equation} where ($v\in R$,$y\in R_{+}$)
\begin{equation}\begin{array}{l}
\theta(r,\tau)=r(2\pi\tau)^{-\frac{1}{2}}\exp(\frac{\pi^{2}}{2\tau})\int_{0}^{\infty}
\exp(-\frac{\xi^{2}}{2\tau}-r\cosh(\xi))\sinh(\xi)\sin(\frac{\pi\xi}{\tau})d\xi.
\end{array}\end{equation}
The $\tau$-integral in the Theorem  can be calculated by an
application of another Yor's formula
\begin{equation}
\int_{0}^{\infty}d\tau\exp(-2\nu\tau)\theta(r,\tau)=I_{\sqrt{2\nu}}(r),
\end{equation} where $I_{\alpha}$ is the modified Bessel function of order
$\alpha$ \cite{grad}. Hence,
\begin{equation}\begin{array}{l}\int_{0}^{\infty}d\tau E[F(B(\tau),A(\tau))]=
\frac{4}{\kappa^{2}}\int F(v,\frac{4}{\kappa^{2}}\vert{\bf p}\vert
y)\exp(2v)\cr
\exp\Big(-\frac{1}{2y}(1+\exp(2v))\Big)I_{2}(y^{-1}\exp(v))y^{-1}dydv
\end{array}\end{equation}
We can apply the formula (91) only to the expectation values (18)
of diffusions without friction (because only in this case the
integrand (18) depends solely on $A$ and $B$) corresponding to the
J\"uttner distribution with $f=0$ (this is the Lorentz invariant
but non-normalizable distribution $dxd{\bf p}\Phi_{E}=dxd{\bf
p}p_{0}^{-1}$). Then, with the proper rescalings we have  (we omit
the index $\mu=2$ in $A$ and $B$)
\begin{equation}
{\bf p}_{\tau}={\bf n}\vert{\bf p}\vert\exp
\Big(2B(\frac{\tau\kappa^{2}}{4})\Big),
\end{equation}
\begin{equation}
{\bf x}_{\tau}={\bf x}-\frac{4}{\kappa^{2}}{\bf n}\vert {\bf
p}\vert A(\frac{\tau\kappa^{2}}{4}).
\end{equation}
We can insert these formulae in eqs.(18) and  (83). After the change
of variables
\begin{equation}
v=\frac{1}{2}\ln(\frac{r}{\vert{\bf p}\vert})
\end{equation}
we obtain from the Theorem (in eq.(18)  the integrals over
$\delta$-function can be performed directly and only the
$v$-integral remains)
\begin{equation}\begin{array}{l}
\Psi_{t}({\bf x},{\bf p})=t^{\prime
-1}\int_{0}^{\infty}dr\frac{r}{\vert {\bf p}\vert}
\exp\Big(-\frac{1}{2t^{\prime}}(\vert {\bf p}\vert +r)\Big) \cr
I_{2}(t^{\prime -1}\sqrt{\vert{\bf p}\vert r} ) \Psi({\bf x}-{\bf
n}ct,{\bf n}r),
\end{array}\end{equation}
where $t^{\prime}=\frac{c\kappa^{2}}{4}t$.

Calculations in the light-cone coordinates for a massive particle
diffusing without a friction  ($f=0$ in eq.(39)) are similar. We
obtain from eq.(58) that $\mu=0$ for $E=0$.Then, from
eqs.(88)-(90) (we change variables $r=p_{+}\exp(2v)$)
\begin{equation}
\begin{array}{l}
\Psi_{t_{-}}(p_{+})=p_{+}\int_{0}^{\infty}d\tau
E\Big[\Psi\Big(p_{+}\exp(2B)\Big)
\exp(2B)\delta(t-\frac{4}{m\kappa^{2}}p_{+}A)\Big]\cr=p_{+}
\int\exp(2v)\Psi(\exp(2v)p_{+})I_{0}(\frac{\exp (v)}{y})
\delta(t-\frac{4}{m\kappa^{2}}p_{+}y)\cr
\exp(-\frac{1}{2y}(1+\exp(2v)))dvy^{-1}dy
=t_{-}^{-1}\int dr
\Psi(r)\exp(-\frac{p_{+}+r}{2t_{-}})I_{0}(\frac{1}{t_{-}}\sqrt{rp_{+}}),
\end{array}\end{equation}
where $t_{-}=\frac{1}{4}\kappa^{2}x_{-}$.

We can generalize this result to a particle in a constant electric
field. Then, from eq.(58) after a rescaling similar to eq.(87) we
obtain
\begin{equation} \mu=\frac{2\alpha E}{\kappa^{2}}.
\end{equation}
Hence, in the transition function we shall have
$I_{\mu}(\frac{1}{t_{-}}\sqrt{p_{+}r})$. The solution of eq.(73)
obtained from eq.(74) is
\begin{equation}
\begin{array}{l}
\Psi_{t_{-}}(p_{+})=t_{-}^{-1}\int dr(\frac{r}{p_{+}})^{\mu}
\Psi(r)\exp(-\frac{p_{+}+r}{2t_{-}})I_{\mu}(\frac{1}{t_{-}}\sqrt{rp_{+}}).
\end{array}\end{equation}

We shall check the results (95)-(98) of the probabilistic averaging
by solving the transport equations (73) and (81) directly.
 Let $\rho=\vert {\bf p}\vert $ then from eq.(78) $\rho$ satisfies the stochastic
 equation
\begin{equation}
d\rho=\frac{3\kappa^{2}c}{2}dt-
 \frac{c^{2}\beta\kappa^{2}}{2}\rho dt+\sqrt{c}\kappa \rho^{\frac{1}{2}}db
\end{equation}
whereas
\begin{equation}
dx^{j}=-c\vert {\bf p}\vert^{-1}p_{j}dt=-cn_{j}dt. \end{equation}
Eq.(100) has the solution \begin{equation} {\bf x}(t)={\bf x}-{\bf
n}ct.
\end{equation}

The transport equation (64) for massless particles with the
J\"uttner friction can be expressed in  the form
\begin{equation}
\partial_{t}\Psi=\frac{c\kappa^{2}}{2} r\partial_{r}^{2}\Psi+(\frac{\sigma}{4}-\frac{\nu}{2}
r)\partial_{r}\Psi -{\bf n}\nabla_{{\bf x}}\Psi
\end{equation}
where $\sigma=6\kappa^{2}c $ and $\nu=\beta c^{2}\kappa^{2}$. The
transport equation (102) is related to the (generalized) Bessel
diffusion
\begin{equation}
\partial_{t}\chi=\frac{\kappa^{2}c}{2}\partial_{\rho}^{2}\chi
+ \frac{\sigma -\kappa^{2} c}{2\rho}\partial_{\rho}\chi-\nu
\rho\partial_{\rho}\chi.
\end{equation}
The diffusion process $\rho_{t}$ solving the diffusion equation
(103) (i.e., $\chi_{t}=E[\chi(\rho_{t})]$) is the solution of the
stochastic equation
\begin{equation}
d\rho=\frac{\sigma -\kappa^{2}c}{2\rho}dt-\nu \rho dt+\kappa db.
\end{equation}
Let \begin{equation}r=\frac{1}{4}\rho^{2} \end{equation}
then
\begin{equation}
dr=\frac{\sigma}{4}dt -\frac{\nu}{2}rdt+\kappa \sqrt{c}\sqrt{r}db.
\end{equation}
If $\nu=0$ then the Kolmogorov transition function ( the probability
density to go from $x$ to $y$ in time $t$) for the Bessel process is
(ref.\cite{ikeda},sec.IV.8; see also \cite{yorpit}\cite{watanabe})
\begin{equation}
P(t,x,y)=\frac{1}{t^{\prime}}(xy)^{1-\frac{\tilde{\sigma}}{2}}y^{\tilde{\sigma}-1}I_{\frac{\tilde{\sigma}}{2}-1}
(\frac{xy}{t^{\prime}})
\end{equation}
where $\tilde{\sigma}=\frac{\sigma}{\kappa^{2}c}$ and
$t^{\prime}=\frac{t\kappa^{2}c}{4}$. Eq.(107) coming from
ref.\cite{ikeda} coincides with our results (95)-(96).

We can calculate the Laplace transform of the process $p_{+}(t)$
 applying the results (91),(96) and (98) of the Theorem of sec.2 to $\Psi=\exp(-\lambda r)$
and using the integrals 6.614 from ref. \cite{grad}
\begin{equation}\begin{array}{l}
E[\exp(-\lambda p_{+}(t_{-},p_{+}))]=t_{-}^{-1}\int dr \exp(-\lambda
r)\exp(-\frac{p_{+}+r}{2t_{-}})I_{0}(\frac{1}{t_{-}}\sqrt{rp_{+}})
\cr=(1+\lambda t_{-})^{-1}\exp\Big(-\lambda p_{+}(1+\lambda
t_{-})^{-1}\Big). \end{array}\end{equation} The diffusion process
(73) in an electric field $E$ has the Laplace transform
\begin{equation}
(1+\lambda t_{-})^{-1-2\alpha E\kappa^{-2}}\exp\Big(-\lambda
p_{+}(1+\lambda t_{-})^{-1}\Big).
\end{equation}The Laplace transform (108)-(109) has been calculated earlier in \cite{ikeda}
(sec.IV.8).

 The solution of the transport equation  with friction
 cannot be calculated with Yor's formula (88).
 The diffusion process (see eqs.(117)-(118) below)
depends on the variables $A$ and $B$ but in eq.(18) we still need
its integral over time. Nevertheless, we can solve the diffusion
equations (72) and (75) directly applying their relation to the
(generalized) Bessel diffusion (104) with $\nu>0$.
 We have  calculated the transition function of the process $\rho_{t}$ with $\nu\neq 0$
 in \cite{habarep}\begin{equation}\begin{array}{l}
 P_{t}(x,y)=\exp(t\nu(\alpha+1))(\sinh \nu t)^{-1}
 y^{\alpha+1}x^{-\alpha}I_{\alpha}(\nu xy(\sinh \nu t)^{-1})
 \cr\exp(-\frac{\nu}{2}(\coth (\nu t)+1)y^{2}-\frac{\nu}{2}(\coth (\nu
 t)-1))x^{2}),
\end{array}\end{equation}
where \begin{displaymath}
\alpha=(\kappa^{2}c)^{-1}\frac{\sigma}{2}-1
\end{displaymath}
($\alpha=2$ in the model of the diffusion of massless particles).
The transition function (108) describes the imaginary time evolution
in quantum mechanics with the potential
$V(x)=\frac{\nu}{2}x^{2}+gx^{-2}$ (then
$\alpha=\frac{1}{2}\sqrt{1+8g}$,\cite{habarep}).

   The Laplace transform of the transition function of
the process $r_{t}$ (with $\nu>0$) starting from $k$ which is the
square of the generalized Bessel process  can be calculated using
the transition function (110) (it has been has been derived earlier
in \cite{ikeda}). We obtain
\begin{equation}\begin{array}{l} E[\exp(-\lambda r_{t})]=\int
dyP_{t}(k,y)\exp(-\lambda
y^{2})=\Big(-\lambda\kappa^{2}\nu^{-1}(\exp(-\frac{\nu}{2}
t)-1)+1\Big)^{-\frac{\sigma}{2c\kappa^{2}}}\cr
\exp\Big(-\Big(-\lambda\kappa^{2}\nu^{-1}(\exp(-\frac{\nu}{2}
t)-1)+1\Big)^{-1}\lambda\exp(-\frac{\nu}{2} t)k\Big).\end{array}
\end{equation}
For the massless particles $\sigma=6c\kappa^{2}$ ,
$\nu=\kappa^{2}c\beta$; for the process $p_{+}$ with the J\"uttner
friction ($f^{\prime}=-1$) we have $\sigma=2c\kappa^{2}$ and
$\nu=\kappa^{2}c\beta$ . The expansion of eq.(111) for the
diffusion of massless particles
\begin{equation}
E[\exp(-\lambda r_{t})]-\frac{1}{2}(c\beta)^{3} \int
dyy^{2}\exp(-c\beta y-\lambda y)
=\sum_{n=1}^{\infty}c_{n}(k)\exp(-n\frac{\nu}{2} t)
\end{equation}
shows that $d\mu=\frac{1}{2} (c\beta)^{3}y^{2}\exp(-c\beta y)$ is
the invariant measure for the process $\rho_{t}$ and that the
convergence to the equilibrium is exponential with the speed
$\frac{2}{\nu}$ proportional to the temperature.

Applying the transition function (110) we can calculate  the
solution of the transport equation (81) as an expectation value over
the stochastic process
\begin{equation}
\Psi_{t}({\bf x},\vert{\bf p}\vert,{\bf n})=E[\Psi({\bf x}-c{\bf
n}t,\rho_{t},{\bf n})] =\int_{0}^{\infty}dyP_{t}(\sqrt{\vert{\bf
p}\vert},\sqrt{y})\psi({\bf x}-c{\bf n}t,y,{\bf n})\end{equation}

From eq.(113) it follows that if $\Psi$ as the function of ${\bf x}$
has no limit at infinity then the limit $t\rightarrow$ does not
exist (this is a consequence of the wave propagation). If the limit
$t\rightarrow \infty$ in eq.(113) exists then it follows  that
$\Psi_{t_{-}}\rightarrow 1$ and $\Phi_{t_{-}}\rightarrow \Phi_{E}$
(up to a normalization constant). Moreover, for the momentum
probability distribution of an "observable" $\phi$ in a "state"
$\Phi$ we obtain

\begin{equation}\begin{array}{l}
lim_{t\rightarrow \infty}\int d{\bf x}d{\bf p}\phi({\bf
p})\Phi_{t}({\bf x},{\bf p}) \cr=\Big(\int d{\bf p}\Phi_{E}({\bf
p})\phi({\bf p})\Big)\frac{1}{8\pi}(\beta c)^{3} \int d{\bf p}d{\bf
x}\exp(-c\beta\vert{\bf p}\vert)\Psi({\bf x},{\bf p})
\end{array}\end{equation}

As an example of the solution (113) we could consider the plane wave
with the initial condition $\Psi({\bf x},\vert{\bf p}\vert,{\bf
n})=\exp(i\vert{\bf p}\vert {\bf nx})$. Then, from eq.(111)
\begin{equation}\begin{array}{l}
\Psi_{t}({\bf x},\vert{\bf p}\vert,{\bf
n})=\Big(\frac{i}{c\beta}({\bf
nx}-ct)(\exp(-\frac{1}{2}\kappa^{2}c\beta t)-1)+1\Big)^{-3}\cr
\exp\Big(i\Big(\frac{i}{c\beta}({\bf
nx}-ct)(\exp(-\frac{1}{2}\kappa^{2}c\beta
t)-1)+1\Big)^{-1}\exp(-\frac{1}{2}\kappa^{2}c\beta t)\vert{\bf
p}\vert({\bf nx}-ct)\Big)\end{array}\end{equation} As a result we
obtain again a wave (note that $\vert{\bf p}\vert({\bf x}-{\bf
n}ct){\bf n}=p_{\mu}x^{\mu}$ in the relativistic notation)  moving
in the direction ${\bf n}$ with a decreasing amplitude and
exponentially decreasing wave vector.

We can solve the stochastic equations (68) and (75) and calculate
expectation values of functions of the  processes with a friction.
The solution of eq.(68) reads
\begin{equation}
\vert{\bf p}_{\tau}\vert=\vert {\bf
p}\vert\exp(\kappa^{2}\tau+\kappa
b_{\tau})\Big(1+\frac{\kappa^{2}}{2}\beta\vert{\bf p}\vert
\int_{0}^{\tau}ds\exp(\kappa^{2}s+\kappa b_{s})\Big)^{-1}
\end{equation}
 The stochastic equation (75)
for $p_{+}$ can also be solved with the result
\begin{equation}
 p_{+}(\tau)= p_{+}\exp(\kappa
b_{\tau})\Big(1+\frac{\kappa^{2}}{2}\beta  p_{+}
\int_{0}^{\tau}ds\exp(\kappa b_{s})\Big)^{-1}
\end{equation}
Unfortunately, with the Yor's formula we are unable to calculate
the expectation value (18). However, we can obtain expectation
values $E[F({\bf p}_{\tau})]$ of any function of the process. This
may be sufficient for a calculation of  some mean values with the
probability distribution $\Phi$ as a solution of the transport
equation (11), for example applying eq.(18)\begin{displaymath}
\int dxd{\bf p}\phi({\bf p})\Phi(x,{\bf p})= \int d{\bf
p}\phi({\bf p})\Phi_{E}({\bf p})\int_{0}^{\infty}d\tau E[\int
d{\bf x}\Psi({\bf x},\hat{{\bf p}}_{\tau}({\bf
p}))\hat{p}_{0}(\tau)].\end{displaymath} The rhs is of the Yor's
form (88) also for the processes (116)-(117) with a friction.
\section{Transport equations in a moving frame} We
consider a covariant form of the equilibrium distribution
\begin{equation} \Phi_{E}=p_{0}^{-1}\exp(f(c\beta^{\mu}p_{\mu})).
\end{equation}
The four-vector $\beta^{\mu}$ can be related to the velocity of
the frame of reference in such a way that in the rest frame
$\beta=(\frac{1}{kT},0,0,0)$, where $T$ is the temperature and $k$
is the Boltzmann constant (see
\cite{cmb}\cite{patria}\cite{kujaw}\cite{weldon}\cite{svet2} for a
discussion of such equilibrium distributions).Then, the
probability distribution $d\sigma(x,p,\beta)=dxd{\bf p}\Phi_{E}$
transforms in a covariant way under Lorentz transformations
$\Lambda$
\begin{displaymath}
 d\sigma(\Lambda x,\Lambda p,\Lambda \beta)=d\sigma(x,p,\beta). \end{displaymath}
An expectation value in a "state" $\Phi$ satisfying the transport
equation (11) is
\begin{equation}\begin{array}{l}
\int d{\bf p}d{\bf x}\phi({\bf x},{\bf p})\Phi_{t}({\bf x},{\bf
p})=\int d\sigma(x,p,\beta)\phi({\bf x},{\bf p})E[\Psi(\hat{{\bf
x}}_{\tau}({\bf x},{\bf p}),\hat{{\bf p}}_{\tau}({\bf p}))].
\end{array}\end{equation}
We applied the formula (18) in order to express the time evolution
of the solution $\Phi_{t}$ of the transport equation by the proper
time. In eq.(119) $\tau(t)$ is expressed by the coordinate time
from the equation $t=(mc)^{-1}\int_{0}^{\tau}\hat{p}_{0}ds$ .

  If $m>0$ then the drift
(31) reads
\begin{equation}
K_{j}=-\frac{m^{2}c^{3}\kappa^{2}}{2}\beta_{j}f^{\prime}+\frac{\kappa^{2}c}{2}\beta^{\mu}p_{\mu}p_{j}f^{\prime}.
\end{equation}
In the massless case
\begin{equation}
K_{j}=\frac{c\kappa^{2}}{2}\beta^{\mu}p_{\mu}p_{j}f^{\prime}.
\end{equation}

The covariant form of the transport equation takes the form
\begin{equation}\begin{array}{l}
\frac{\partial}{\partial
t}\Psi=\frac{c}{p_{0}}p_{j}\frac{\partial}{\partial x_{j}}\Psi

+\frac{\kappa^{2}m^{3}c^{3}}{2p_{0}}(\delta_{jk}+m^{-2}c^{-2}p_{j}p_{k})\frac{\partial}{\partial
p_{j}}\frac{\partial}{\partial
p_{k}}\Psi\cr+\frac{m\kappa^{2}c}{p_{0}}(\frac{3}{2}+\frac{1}{2}
c\beta^{\mu}p_{\mu}f^{\prime} )p_{k}\frac{\partial}{\partial
p_{k}}\Psi+\frac{m^{3}c^{4}\kappa^{2}}{2p_{0}}\beta_{k}\frac{\partial}{\partial
p_{k}}\Psi.
\end{array}\end{equation}
In the massless case
 \begin{equation}\begin{array}{l}
\frac{\partial}{\partial
t}\Psi=\frac{c}{p_{0}}p_{j}\frac{\partial}{\partial x_{j}}\Psi

+\frac{c\kappa^{2}}{2p_{0}}p_{j}p_{k}\frac{\partial}{\partial
p_{j}}\frac{\partial}{\partial p_{k}}\Psi+\frac{\kappa^{2}}{2p_{0}}
c^{2} \beta^{\mu}p_{\mu}p_{k}f^{\prime} \frac{\partial}{\partial
p_{k}}\Psi+\frac{c\kappa^{2}}{2p_{0}} p_{k}\frac{\partial}{\partial
p_{k}}\Psi.
\end{array}\end{equation}
In the proper time formalism the diffusion process solving the
diffusion equation (15) satisfies the equation
\begin{equation}
d\hat{p}_{j}=\frac{3\kappa^{2}}{2}\hat{p}_{j}d\tau-\frac{m^{2}c^{3}\kappa^{2}}{2}\beta_{j}f^{\prime}d\tau
+\frac{\kappa^{2}}{2}\beta^{\mu}\hat{p}_{\mu}\hat{p}_{j}f^{\prime}
d\tau +\kappa e^{j}_{n}db^{n}.
\end{equation}
where $e^{j}_{n}$ (the square root of the metric in the second order
differential operator (21), see \cite{habapre}) has been defined in
eq.(69). The diffusion process $\hat{{\bf p}}_{\tau}$ (124) is the
same as the process ${\bf p}_{\tau}$ generated by ${\cal A}$
(eq.(7)) because ${\hat{\cal A}}=\Phi_{E}^{-1}{\cal A}\Phi_{E}={\cal
A}$. We have obtained this equality by direct calculations but there
should be a deeper reason for it.

 We can derive from eq.(124) the formula for
$dp_{0}$. Let us define
\begin{equation}
\pi=\beta^{\mu}p_{\mu}
\end{equation}
and
\begin{equation}
X=\beta^{\mu}x_{\mu}.
\end{equation}
Then, a direct calculation (using the Ito stochastic calculus
\cite{ikeda}) leads to the proper time stochastic equations

\begin{equation}\begin{array}{l}
d\pi=\frac{3\kappa^{2}}{2}\pi d\tau
+\frac{c\kappa^{2}}{2}\pi^{2}f^{\prime}d\tau\cr
-\frac{m^{2}c^{3}\kappa^{2}}{2}\beta^{\mu}\beta_{\mu}f^{\prime}d\tau
+\frac{\kappa}{p_{0}}(\beta_{0}p_{j}-\beta_{j}p_{0})e^{j}_{n}db^{n}
\end{array}\end{equation}
\begin{equation}
\frac{dX}{d\tau}=\pi.
\end{equation}
In the massless case eq.(127) reads
\begin{equation}\begin{array}{l}
d\pi=\frac{3\kappa^{2}}{2}\pi d\tau
+\frac{c\kappa^{2}}{2}\pi^{2}f^{\prime}d\tau +\kappa\pi db.
\end{array}\end{equation}

Using the variables $(X,\pi)$ we can write a proper time diffusion
equation for an evolution of the probability distribution (13)
$\Psi(X,\pi)$ assuming that it depends solely on the variables
$(X,\pi)$
\begin{equation}\begin{array}{l}
\partial_{\tau}\Psi=\pi\partial_{X}\Psi+(\frac{3\kappa^{2}}{2}\pi
+\frac{c\kappa^{2}}{2}\pi^{2}f^{\prime}
-\frac{m^{2}c^{3}\kappa^{2}}{2}\beta^{\mu}\beta_{\mu})\partial_{\pi}\Psi
\cr
+\frac{\kappa^{2}}{2}(\pi^{2}-m^{2}c^{2}\beta^{\mu}\beta_{\mu})\partial_{\pi}^{2}\Psi
\end{array}\end{equation}
In the massless case this diffusion equation reads
\begin{equation}\begin{array}{l}
\partial_{\tau}\Psi=\pi\partial_{X}\Psi+(\frac{3\kappa^{2}}{2}\pi
+\frac{c\kappa^{2}}{2}\pi^{2}f^{\prime})\partial_{\pi}\Psi
+\frac{\kappa^{2}}{2}\pi^{2}\partial_{\pi}^{2}\Psi
\end{array}\end{equation}
The transport equation for massless particles is the same as the one
for the Bessel diffusion
\begin{equation}
\partial_{X}\Psi=\frac{\kappa^{2}}{2}\pi\partial^{2}_{\pi}\Psi
+(\frac{3\kappa^{2}}{2} +\frac{c\kappa^{2}}{2}\pi
f^{\prime})\partial_{\pi}\Psi.
\end{equation}
In order to preserve an interpretation of $\beta$ as the temperature
the four-vector $\beta$ should be time-like. We can see from
eq.(130) that the model simplifies substantially (the transport
equation reduces to the Bessel diffusion) if we let
$\beta^{\mu}\beta_{\mu}\rightarrow 0$. In a formal limit
$\beta=(\beta_{-},0,0,0)$ (when $\beta$ is on the light cone) and
with $\partial_{\tau}\Psi=0$ we obtain the model (72) corresponding
to the light cone coordinates with the exception of the factor
$\frac{3\kappa^{2}}{2}\partial_{\pi}$ in eq.(130) which is replaced
by $\frac{\kappa^{2}}{2}\partial_{+}$ in eq.(72). The incorrect
factor in the  formal limit comes from the incorrect probability
measure in this limit which in the light-cone coordinates should be
 $dp_{1}dp_{2}dp_{+}p_{+}^{-1}\exp f$ and not $d{\bf
p}p_{0}^{-1}\exp f$. The factor $p_{+}^{-1}$ is responsible for the
difference between the term $\frac{3\kappa^{2}}{2}\partial_{\pi}$ in
eq.(130) and $\frac{\kappa^{2}}{2}\partial_{+}$ in eq.(72).

 Relativistic invariance means that a Lorentz transformation of the processes
 $X$ and $\pi$ can be shifted into a transformation of the frame of
 reference, i.e., for a function $F$

\begin{equation}
E[F(\Lambda x(\tau,\beta,\Lambda x, \Lambda p),\Lambda
p(\tau,\beta,\Lambda p))]= E[F(x(\tau,\Lambda^{-1}\beta, x,
p),p(\tau,\Lambda^{-1}\beta, p))]\end{equation}

\section{Discussion}
In our earlier papers we have developed a formalism of relativistic
diffusions evolving in the proper time. Such a formalism is useful
because the explicit relativistic invariance is a strong guiding
principle  when building  relativistic diffusion equations. In this
paper we have shown that a solution of the stochastic equations
evolving in the proper time allows a construction of the solution of
the transport equation (in the laboratory time) by means of an
integration over the proper time. We have shown how the method works
in some models which can be solved exactly. These examples are not
typical for the proper time equations. For massless particles the
proper time is just an affine parameter on the trajectory without a
physical meaning. A function depending solely on $p_{+}$ can
describe a massive particle whose motion is restricted to a straight
line (in the direction of the third axis). The results concerning
the transport equations described by the Bessel diffusion of momenta
could find applications in astrophysics, in plasma physics and in
heavy ion collisions. The momentum  $\vert{\bf p}\vert$ evolves in
time as the Bessel process which has some distinguished features
among all diffusion processes \cite{yorpit}\cite{watanabe}. In
particular, the
 (generalized) Bessel diffusion has the exceptional property that
an exponential initial distribution remains exponential for any
time.

 In general,
the method of proper time could be applied for approximate
calculations and computer simulations. If $m>0$ then an expansion
parameter $(mc)^{-1} $ appears in all our equations. We could use it
in the Theorem (eq.(18)) in order to calculate the integral over
$\tau$ with an expanded argument of the $\delta $-function
($\hat{{\bf p}}_{s}={\bf b}_{s}$ in the lowest order of the
expansion in $(mc)^{-1}$)\begin{displaymath}
\delta\Big(t-\tau-\frac{1}{2m^{2}c^{2}}\int_{0}^{\tau}ds {\bf
b}_{s}^{2}\Big).
\end{displaymath}
We can now calculate the expectation value (18) in a way similar to
the one in this paper  because the probability distribution of the
integral  $\int_{0}^{\tau}ds {\bf b}_{s}^{2}$ is known.

The original relativistic diffusion  in the proper time of Schay
\cite{schay} and Dudley \cite{dudley}  is explicitly Lorentz
invariant because its proper time evolution is generated by a
Lorentz invariant differential operator. In applications we need
diffusions which have a limit for a large time. The necessary drag
terms which force the process to an equilibrium cannot be Lorentz
invariant. The reason is that the notion of the equilibrium is
itself frame dependent. We described this frame dependence in the
last section. The stochastic process transforms in a covariant way
with respect to the Lorentz transformation if together with the
process we transform also the frame.


\begin{thebibliography}{99}\bibitem{hanggirev}J. Dunkel and P. H\"anggi,Phys.Rep.{\bf 471},1(2009)

\bibitem{debbaschrev}C. Chevalier and F. Debbasch, AIP Conf.Proc.{\bf
913},42(2007)
\bibitem{chev}C. Chevalier and F. Debbasch,

Journ.Math.Phys.{\bf 49},043303(2008)


\bibitem{schay} G.Schay,PhD thesis,Princeton University,1961
\bibitem{dudley} R.Dudley, Arkiv for Matematik,{\bf 6},241(1965)
\bibitem{lejan}J. Franchi and Y. Le Jan, Commun.Pure Appl.Math.{\bf 60},187(2007)
\bibitem{franchi}J. Franchi, Commun.Math.Phys.{\bf 290},523(2009)
\bibitem{habajpa}Z. Haba,Journ.Phys.{\bf A42},445401(2009)
\bibitem{habagen}Z. Haba, arXiv:0909.2880
 \bibitem{habapre}Z. Haba, Phys.Rev.{\bf E79},021128(2009)

\bibitem{habakomp}Z.Haba,Mod.Phys.Lett.A, arXiv:0910.2253
\bibitem{komp}A.S. Kompaneets, JETP,{\bf 47},1939(1956)(in Russian)




\bibitem{rybicki}G.B. Rybicki and A.P. Lightman,

Radiative Processes in Astrophysics,Wiley-VCH,1979
\bibitem{peebles}P.J.E. Peebles, Physical Cosmology, Princeton
University Press,1971
\bibitem{dodelson}S. Dodelson, Modern Cosmology, Academic Press, New
York,2003
\bibitem{ion}R.Rapp and H. van Hees,Int.Journ.Mod.Phys.{\bf E}(2009) arXiv:0903.1096
\bibitem{svet} B.Svetitsky, Phys.Rev.{\bf D37},2484(1988)
\bibitem{stewart}J.M. Stewart, Non-equilibrium Relativistic
Kinetic Theory,Lect.Notes in Physics,Vol.10,Springer,1971
\bibitem{israel}W. Israel, Journ.Math.Phys.{\bf 4},1163(1963)

\bibitem{zachar} P. Carruthers and F. Zachariasen,

Phys.Rev.{\bf D13},950(1976)
\bibitem{elze} H.-Th. Elze and U.Heinz,
Phys.Rep.{\bf 183},81(1989)
\bibitem{kelly}
P.F. Kelly, Q. Liu, G.Lucchesi and C. Manuel,

Phys.Rev.Lett.{\bf 72},3461(1994),Phys.Rev.{\bf D50},4209(1994)

\bibitem{landau} L.D. Landau and E.M. Lifshits,

 Field Theory, Pergamon Press,Oxford,1981
\bibitem{skorohod}I.I. Gikhman and A.V.
Skorohod,

Stochastic Differential Equations,Springer,Berlin,1972
\bibitem{feynman}R.P. Feynman, Phys.Rev.{\bf 80},440(1950),{\bf
84},108(1951)


\bibitem{yor}
H. Matsumoto and M. Yor, Probability Surveys,{\bf 2},312(2005)
\bibitem{yor2}M.Yor,Adv.Appl.Prob.{\bf 24},509(1992)
\bibitem{dun}J. Dunkel, P. H\"anggi and S. Weber, Phys.Rev.{\bf E
79},010101(R)(2009)
\bibitem{ikeda} N. Ikeda and S.
Watanabe, Stochastic Differential Equations and Diffusion
Processes,North Holland,1981
\bibitem{alili}L. Alili, H. Matsumoto and T. Shiraishi, in Lecture
Notes in Math., Vol.1755,Springer,2001


\bibitem{yorpit}J.Pitman and M.Yor,
Z.Wahr.verw.Geb.{\bf 59},425(1982)

\bibitem{watanabe}T. Shiga and S.
Watanabe, Z.Wahr.verw.Geb.{\bf 27},37(1973)

\bibitem{grad}I.S.Gradshteyn , I.M. Ryzhik, Tables of Integrals,Series and Products,


Nauka, Moscow,1971(in Russian)





























\bibitem{juttner}F. J\"uttner, Ann.Phys.(Leipzig){\bf 34},856(1911)


\bibitem{habarep}Z. Haba, Reports Math.Phys.{\bf 18},257(1980)


\bibitem{cmb}P.J.E. Peebles and D.T. Wilkinson, Phys.Rev.{\bf 174},2168(1968)

G.R. Henry, R.B. Feduniak, J.E. Silver and M.A. Peterson,

Phys.Rev.{\bf 176},1451(1968)
\bibitem{patria}R.K. Patria,Proc.Phys.Soc.{\bf 88},791(1966)


\bibitem{kujaw}J.H. Eberly and A.Kujawski, Phys.Rev.{\bf
155},10(1967)

\bibitem{weldon}H.A. Weldon, Phys.Rev.{\bf D26},1394(1982)


\bibitem{svet2}T.Matsui,B. Svetitsky and
L.D.McLerran,Phys.Rev.{\bf D34},783(1986)






\end{thebibliography}
\end{document}